\RequirePackage{fix-cm}
\documentclass[smallextended]{svjour3}       
\smartqed  
\usepackage{graphicx}
\usepackage{amssymb}
\usepackage{amsmath}
\usepackage{graphicx}
\usepackage{dcolumn}
\usepackage{bm}

\begin{document}
\title{Total quantum coherence and its applications\thanks{This work was supported by the National Natural Science Foundation of China,
under Grant No.11375036, the Xinghai Scholar Cultivation Plan
and the Fundamental Research Funds for the Central Universities under Grant
No. DUT15LK35 and No. DUT15TD47.
}}
\author{Chang-shui Yu         \and
        Si-ren Yang \and Bao-qing Guo
}


\institute{Chang-shui Yu\at
              School of Physics and Optoelectronic Technology, Dalian University of
Technology, Dalian 116024, P. R. China \\
                         \email{quaninformation@sina.com; or ycs@dlut.edu.cn}             \\
\and Si-ren Yang\at School of Physics and Optoelectronic Technology, Dalian University of
Technology, Dalian 116024, P. R. China \\
           \and
           Bao-qing Guo \at
              School of Physics and Optoelectronic Technology, Dalian University of
Technology, Dalian 116024, P. R. China 
}

\date{Received: date / Accepted: date}

\maketitle

\begin{abstract} Quantum coherence is the most fundamental feature of quantum mechanics. The usual understanding of it depends on the choice of the basis, that is, the coherence of the same quantum state is different within different reference  framework. To reveal all the potential coherence, we present the total quantum coherence measures in terms of two different methods. One is  optimizing maximal basis-dependent coherence with all potential bases considered  and  the other is quantifying the distance between the state and the  incoherent state set.   Interestingly, the coherence measures based on relative entropy and $l_2$ norm have the same form in the two different methods. In particular, we show that the measures based on the non-contractive $l_2$ norm is also a good measure different from the basis-dependent coherence.   In addition, we show that all the measures are
analytically calculable and have all the good properties. The
experimental schemes for the detection of these coherence measures are also
proposed by multiple copies of quantum states instead of reconstructing the full density matrix. By studying one type of quantum probing schemes, we find that both the normalized trace in the scheme of deterministic quantum computation with one qubit (DQC1) and the overlap of two states in quantum overlap measurement schemes (QOM) can be well  described by the change of total coherence of the probing qubit. Hence the nontrivial probing always leads to the change of the total coherence.
 \end{abstract}

\section{Introduction}

Coherence is not only the essence of the interference phenomena but also the
foundation of quantum theory [1]. It is almost directly or indirectly
related to all the intriguing quantum phenomena. The most remarkable
phenomena are quantum correlation including quantum entanglement. Both can
be understood as the combination of the coherence and the tensor product
structure of state space and play important roles in quantum information
processing tasks (QIPTs) [2-5]. In addition, it is also shown that quantum coherence has been widely applied in quantum thermal engine [6,7],  biological system [8] and quantum
parallelism [9].

In quantum information, quantum feature such as  entanglement [10] and quantum correlation [11], due to the potential application in QIPTs,  can be well quantified from
the resource theory point of view.  Recently,  In
the same manner a rigorous framework has been developed for the
quantification of quantum coherence [12]. It points out that a good
coherence measure should satisfy three conditions: 1) The incoherent states
have no coherence; 2) (Monotonicity) Incoherent completely positive and
trace preserving maps cannot increase the coherence, or the average
coherence is not increased under selective measurements; 3) (Convexity) It
is not increased under the mixing of quantum states. Meanwhile it also presented several
good coherence measures. However, such
coherence measures strongly depend on the choice of the basis. This means that a quantum state
can have certain coherence in one basis, but it could possess more, less, or none coherence in the other basis. Even though such a basis-dependent quantification of quantum coherence is consistent with our intuitive understanding (such as the contribution of off diagonal entries of density matrix), this could only consider the partial contribution of coherence of a state, once one is allowed to freely select the basis. In particular,   the
change of basis is a quite easy thing or at a small price at practical scenarios. Taking the linear optics for an
example, one can only rotate the wave plate to get to another framework [13]. Since quantum coherence can be understood as
the useful resource, why not try one's best to extract it as many as possible? So it is natural to consider, with all potential basis taken into account, how much coherence a state possesses or what is the maximal coherence in a state.

In this paper, we present the total coherence measure to quantify all the
contributions of the quantum coherence in a state. The most distinct feature
of this measure is that it only covers the property of a state instead of
the external observable (the choice of basis). We give several analytically
calculable coherence measures in two different frameworks: optimization among all potential bases or quantifying the distance between the state and the incoherent state set. We find that all measures satisfy the mentioned three
properties. In particular, one can find that
the measure based on $l_2$ norm is also a valid candidate, even though the $%
l_2$ norm is not contractive. In addition, we  find that the coherence measures based on relative entropy and the $l_2$ norm have the same result in the different frameworks. From the angle of the experimental detection,
we give an explicit scheme to physically detect these measures. It is
shown that such detections do not require reconstructing the full density
matrix. As an application, we study the total coherence in the DQC1-like
quantum probing schemes [14,15] including the QOM [16]. As we know,
DQC1-like quantum schemes show quantum speedup, but what the source of the
speedup is remains open. Here instead of finding the exact source, we study
what cost is needed to pay for such schemes. It is found that both the
normalized trace in DQC1 and the overlap of two states in QOM can be well
described by the change of the total coherence of the probing qubit. In
other words, the nontrivial quantum probing always gives rise to the change
of the total coherence. The paper is organized as follows. We first propose
various total coherence measures; then we present the properties of these
measures; and then we study the total coherence in the DQC1-like quantum
probing schemes; Finally, we give a summary and discussions.

\section{The total coherence measure}

The classical coherence is usually characterized by the frequencies and the
phases of different waves, but a good definition of quantum coherence
stemming from the superposition of state (a single wave) depends not only on
the state itself but also on the associated observable. The physical root of
such a definition is that the measurement on the observable can reveal the
interference pattern provided that the observable does not commute with the
considered density matrix. In this sense, it is obvious that the coherence
measure will have to depend on the framework (or basis) that the density matrix
is given in [17,18]. Therefore, there are naturally two ways to
quantifying quantum coherence: one is based on the commutation, the other is
based on the distance. Based on the former, Ref. [19] used the skew
information to measure the coherence, and based on the latter, Ref. [12]
proposed several measures. We also used $l_{1}$ norm to study the source of
quantum entanglement [20]. Considering the potential classification of
coherence of composite quantum system, we have provided a new angle to
understand the geometric quantum discord, quantum non-locality and the
monogamy of coherence [21]. Here we shall consider the maximal coherence
with different bases taken into account, or the total coherence which a state
could have. Therefore, a natural method to doing so is to maximize the basis-dependent coherence by taking into account all the potential bases. In addition,  as mentioned before, the coherence can be embodied by the commutation between the state and some particular observable. Since we consider all
potential bases or (observables), it is implied that the incoherent state
requires that the density matrix should commute with all observables. The
direct conclusion is that the incoherent state is the maximally mixed state $%
\frac{\mathbf{1}_{n}}{n}$ with $n$ denoting the dimension of the state and $%
\mathbf{1}_{n}$ denoting the $n$-dimensional unity. So one can easily
construct the coherence measure based on the 'distance'  [9].
In the following, we will consider the coherence measures both by optimizing the basis and by the distance.
 
\textbf{Coherence based on basis optimization.}-With the different bases
considered, we can define the total coherence based on the optimization of
basis as follows.%
\begin{equation}
C\left( \rho \right) =\max_{U}\left\Vert U\rho U^{\dag }-\sigma
_{U}\right\Vert ,
\end{equation}%
where $\left( \sigma _{U}\right) _{ii}=\left( U\rho U^{\dag }\right) _{ii}$
denotes the diagonal matrix and $\left\Vert \cdot \right\Vert $ denotes some
good norms or distance functions. For example, we can employ the $l_{1}$
norm, $l_{2}$ norm, relative entropy and so on. One can also use the trace
norm and Fidelity, but the incoherent state $\sigma _{U}$ is usually not given
by $\left( U\rho U^{\dag }\right) _{ii}$, but some particular states in the
incoherent set [12]. The skew information can also be employed, but no explicit incoherent state is required. In order to provide
an explicit expression of the total coherence, next we will list some
coherence measures by the particular choice of the "norms". 

(1)  $l_{2}$ norm could be the most easily calculable norm. But it is not contractive, so in many cases an unphysical result could appear [12]. In the current case, one will find in the paper that $l_{2}$ norm can be safely used to quantify the total coherence measure. Based on $l_{2}$ norm, we have 
\begin{eqnarray}
C_{2}\left( \rho \right) &=&\max_{U}\left\Vert U\rho U^{\dag }-\sigma
_{U}\right\Vert _{2}  \nonumber \\
&=&Tr\rho ^{2}-\min_{\left\vert i\right\rangle
}\sum\limits_{i=1}^{n}\left\vert \left\langle i\right\vert \rho \left\vert
i\right\rangle \right\vert ^{2}  \nonumber \\
&=&Tr\rho ^{2}-\frac{1}{n}.
\end{eqnarray}%
The minimum can  be reached because there always exists  the basis $\{\left\vert i\right\rangle \}$ such that
the diagonal entries of $\rho $ are uniform.

 (2) If we employ the relative entropy [9], the total coherence can be given by 
\begin{eqnarray}
C_{re}\left( \rho \right) &=&\max_{U}S\left( U\rho U^{\dag }||\sigma
_{U}\right)  \nonumber \\
&=&\text{Tr}\rho \log \rho -\min_{U}\sum\limits_{i=1}^{n}\sigma _{U}\log
\sigma _{U}  \nonumber \\
&=&\log n-S\left( \rho \right) ,
\end{eqnarray}%
with $S\left( \rho ||\sigma \right) =Tr\rho \log \rho -Tr\rho \log
\sigma $ denoting the relative entropy of $\rho$ and $\sigma$. The minimum is also achieved by the basis subject to the uniform
distribution of the diagonal entries of $\rho $. 

(3) Based on skew information,
we will have a different definition. The skew information [22-24] for a
density matrix $\rho $ and an observable $K$ is given by $I\left( \rho
,K\right) =-\frac{1}{2}$Tr$\left[ \sqrt{\rho },K\right] ^{2}$. Based on the
skew information, the total coherence can be defined by%
\begin{eqnarray}
C_{I} &=&\max_{\{\left\vert k\right\rangle \}}\sum\limits_{k=1}^{n}\left(
\left\langle k\right\vert \rho \left\vert k\right\rangle -\left\langle
k\right\vert \sqrt{\rho }\left\vert k\right\rangle ^{2}\right)  \nonumber \\
&=&1-\min_{\{\left\vert k\right\rangle \}}\left\langle k\right\vert \sqrt{%
\rho }\left\vert k\right\rangle ^{2}  \nonumber \\
&=&1-\frac{1}{n}\left( \sum \sqrt{\lambda _{j}}\right) ^{2},
\end{eqnarray}%
where $\lambda _{i}$ is the eigenvalue of $\rho $. In particular, one can
find that the minimum can always be reached when in the basis $\left\{
\left\vert k\right\rangle \right\} $ the diagonal entries of $\sqrt{\rho }$
are uniform. Why $C_{I}$ can quantify the coherence can be easily found as
follows. Given a basis $\left\{ \left\vert k\right\rangle \right\} $, $\rho $
can be diagonalized by the basis $\left\{ \left\vert k\right\rangle \right\} 
$, iff $\left[ \sqrt{\rho },\left\vert k\right\rangle \left\langle
k\right\vert \right] =0$ for all $\left\vert k\right\rangle $. So the skew
information in the basis $\left\{ \left\vert k\right\rangle \right\} $ can
quantify the coherence of $\rho $ in this basis. Considering all the
potential basis, the maximal  $C_{I}$ naturally quantifies the total
coherence as mentioned above. This definition should be distinguished from
that in Ref. [19] where the coherence could depend on the eigenvalues of the
observable. 

 $l_{1}$ norm of a matrix is defined by the sum of the absolute values of all the entries of the matrix. It is also a good norm even though it is not a unitary-invariant norm. With this norm, the total coherence can be given by $%
C_{1}\left( \rho \right) =\max_{U}\sum\limits_{i\neq j}\left\vert
\left\langle i\right\vert U\rho U^{\dag }\left\vert j\right\rangle
\right\vert $, with the maximum reached when all the diagonal entries
of $U\rho U^{\dag }$ equal $\frac{1}{n}$. However, because $l_{1}$ norm is
not unitary-invariant, the optimal result of $C_{1}\left( \rho \right) $ for a general $\rho$ (especially in high dimensional Hilbert space)
cannot be easily given. But one can find that $C_1$ is unitary- invariant, because the optimization compensates for it. In addition, the explicit expressions of the
total coherence based on trace norm and the Fidelity can not be easily given
because the nearest incoherent state can not be determined in general cases.

\textbf{Coherence based on distance.}-Since the completely incoherent
state is $\frac{\mathbf{1}_{n}}{n}$, one can always define the total
coherence based on the distance between a given state and $\frac{\mathbf{1}%
_{n}}{n}$. Using some (unitary-invariant) norms or distance functions $\left\Vert\cdot\right\Vert$, we have
\begin{equation}
\tilde{C}\left( \rho \right) =\left\Vert \rho -\frac{\mathbf{1}_{n}}{n}%
\right\Vert . 
\end{equation}
Since no optimization is included, all the coherence measures can be easily calculated so long as one
selects a proper function $\left\Vert\cdot\right\Vert$. For example, one can easily find
the explicit form of the total coherence measure $\tilde{C}\left( \rho
\right) $ based on trace norm, Fidelity and so on. Here, we would
like to emphasize the following several candidates. 

($\tilde{1}$) If the  relative entropy is used, one can find 
\begin{eqnarray}
\tilde{C}_{re}\left( \rho \right) &=&\text{Tr}\rho \log \rho -\text{Tr}\rho
\log \frac{\mathbf{1}_{n}}{n}  \nonumber \\
&=&\log n-S\left( \rho \right) =C_{re}\left( \rho \right) .
\end{eqnarray}%

($\tilde{2}$) If l$_{2}$ norm is selected, we have 
\begin{eqnarray}
\tilde{C}_{2}\left( \rho \right) &=&\left\Vert \rho -\frac{\mathbf{1}_{n}}{n}%
\right\Vert _{2}  \nonumber \\
&=&Tr\rho ^{2}-\frac{1}{n}=C_{2}\left( \rho \right) .
\end{eqnarray}

It is obvious that the total coherence measures based on the relative entropy and the $l_2$ norm have the same final expressions in the different frameworks. Because $l_{1}$ norm could be changed by a unitary operation, the coherence based on it has to take some optimization on the unitary transformations, that is, $\tilde{C_1}\left( \rho \right) =\max_U\left\Vert U\rho U^\dagger-\frac{\mathbf{1}_{n}}{n}%
\right\Vert _1$.
Here we would like to emphasize that our coherence measures actually are closely related to the purity. One knows that the purity of a state $\rho$ is defined by $P(\rho)=Tr\rho^2$. It is obvious that $P=1$ for pure states and $\frac{1}{n}\leq P<1$ for mixed state, based on which one can design many other similar quantities for purity such as $1-S(\rho)$ and $(\sum_i\sqrt{\lambda_i})^n$ with $n\geq 1$ and so on. These purities reach maximum value for pure state and nonzero minimum value for maximally mixed states. Thus the presented total coherence can be regarded as a displacement on the purity. In this sense, we give the purity a new understanding by the coherence and \textit{vice versa}. 

One should note that our coherence measures are defined different from the basis-dependent coherence, so the criteria for a good measure should be different either. Next, we will list the useful properties that these measures satisfy, meanwhile they could form new criteria for a basis-independent coherence measure. 

\section{Properties}

In what follows, we will list several good properties that our above  total coherence measures satisfy. In particular, we will show that the coherence measure based on $l_2$ norm is still a monotone, even though it is not a contractive norm. This could provide great convenience for the future applications.

(I) \textit{Maximal for pure states and vanishing for incoherent states.-}
It is easy to find that all the coherence measures vanish for maximally
mixed state $\frac{\mathbf{1}_{n}}{n}$ and arrive at its maximal value for
pure states. This can be well understood, since any pure state can be
converted to a maximally coherent state by changing basis.

(II)\textit{Invariant under unitary operations.-}The most obvious feature,
based on the definitions, is that all these measures are invariant under
unitary transformations.

(III)\textit{Convexity.-} All the measures are convex. That is, the total
coherence will not increase under mixing. This can be found from the fact
that all the norms satisfy the triangle inequality. For the squared $l_2$
norm, one also needs to consider the convexity of the quadratic function. In
addition, we know that the fidelity is strongly concave, the von Neumann
entropy is concave and the skew information is convex, so this property can be 
easily proved.

(IV)\textit{Monotonicity.-} This property will have different contents from that for the basis-dependent coherence measure. Just as in entanglement theory and coherence
measure [12], the definitions of entanglement monotone and coherence
monotone require the non-entangling operations and the basis-dependent
incoherent operations, respectively. Now let the incoherent operation be
given in Kraus representation as $\$_{K}=\left\{ \left. K_{n}\right\vert
\sum K_{n}^{\dagger }K_{n}=\mathbf{1}\right\} $. If we follow the rules of
non-entangling operations and the basis-dependent incoherent operations the
elements of which cannot generate entanglement or coherence, one can easily
find that $K_{n}$ should be a unitary transformation neglecting a constant
(corresponding to probability). Therefore, here the incoherent operations can be
rewritten as $\$_{U}=\left\{ \left. U_{n}\right\vert \sum
p_{n}U_{n}^{\dagger }U_{n}=\mathbf{1,}U_{n}^{\dagger }U_{n}=\mathbf{1}%
\right\} $. A simple algebra can show that the average total coherence
equals the total coherence of the original state before the operation, but
the total coherence of the final state after the operation will not be
increased due to the convexity. We would like to emphasize that the coherence measure based on $l_2$ norm also satisfies this property, so it can be used safely.

(V) \textit{Coherence doesn't increase under the special  POVM.-}
This is another interesting property for the total coherence. Let's consider such
an operation that is given in Kraus representation as $\$_{I}=\left\{ \left.
K_{n}\right\vert \sum K_{n}^{\dagger }K_{n}=\sum K_{n}K_{n}^{\dagger }=%
\mathbf{1}\right\} $. One can find that this operation can not create any
coherence from the incoherent state $\frac{\mathbf{1}_{n}}{n}$, even though
the single element such as $K_{n}$ may produce coherence. Therefore, the
average total coherence could be increased by $\$_{I}$. However, it is
interesting that the total coherence of the final state (the final ensemble
generated by $\$_{I}$ on the original state) after this operation can not be
increased. This conclusion may be drawn from the fact that all the above
employed quantifications but $l_{2}$ norm are contractive. However, one can
also prove that this property is satisfied for the case of $l_{2}$ norm. The
proof is given as follows.

Let $\rho $ denote the density matrix that we want to consider. The final
state after the operation can be given by $\tilde{\rho} =\sum_{ij}K_{i}\rho
K_{i}^{\dag }$. Thus $Tr \tilde{\rho} ^{2}=\sum_{ij}K_{i}\rho K_{i}^{\dag
}K_{j}\rho K_{j}^{\dag }=Tr\sum_{m}\Lambda A_{m}\Lambda A_{m}^{\dag }$,
where we use the eigenvalue decomposition of $\rho =U\Lambda U^{\dag }$ with 
$\Lambda $ denoting the diagonal matrix of eigenvalues $\lambda _{i}$ and $%
A_{m}=U^{\dag }K_{i}^{\dag }K_{j}U$ with $m=(ij)$. Since $\sum
K_{n}^{\dagger }K_{n}=\sum K_{n}K_{n}^{\dagger }=\mathbf{1}$, it is obvious
that $\{A_{m}\}$ defines a Positive Operator-Valued Measurement (POVM) and $%
\sum_{m}A_{m}^{\dag }A_{m}=\sum_{m}A_{m}A_{m}^{\dag }=\mathbf{1}$ which
implies $\sum_{jm}\left\vert [A_{m}]_{ij}\right\vert
^{2}=\sum_{im}\left\vert [A_{m}]_{ij}\right\vert ^{2}=1$. Expand $Tr \tilde{%
\rho} ^{2}$, we will have 
\begin{eqnarray}
Tr \tilde{\rho} ^{2}&= &\sum_{ijm}\left\vert [A_{m}]_{ij}\right\vert
^{2}\lambda _{i}\lambda _{j}  \nonumber \\
&\leq& \left[ \sum\limits_{i}\lambda _{i}^{2}\right] ^{1/2}\left[
\sum\limits_{i}\left( \sum_{jm}\left\vert [A_{m}]_{ij}\right\vert
^{2}\lambda _{j}\right) ^{2}\right] ^{1/2}  \nonumber \\
&\leq& \left[ \sum\limits_{i}\lambda _{i}^{2}\right] ^{1/2}\left[
\sum\limits_{ij}\sum_{m}\left\vert [A_{m}]_{ij}\right\vert ^{2}\lambda
_{j}^{2}\right] ^{1/2}  \nonumber \\
&=&\sum\limits_{i}\lambda _{i}^{2}=Tr\rho^{2}.
\end{eqnarray}
This shows that the total coherence based on $l_{2}$ norm is not increased
under the operation $\$_{I}$ either.

\section{Measurable total coherence}

In the above sections, we mainly consider the mathematical approaches to
measuring the total coherence. How can we directly measure the coherence
experimentally? In fact, one can note that  the presented
measures  can be expressed by the function of the eigenvalues of the density matrix $\rho $%
, for example, Eqs. (2,3,4,6,7). Since the eigenvalues of a density can be directly measured ( for example, the schemes for the measurable entanglement and discord [25-27]), our presented coherence can be naturally determined. However, for integrity and the latter use, we would like to briefly describe the concrete implementation. Since $%
Tr\rho ^{n}=\sum_{i=1}^{N}\lambda _{i}^{n}$ for any $N$-dimensional density
matrix $\rho $, one can only set $n=1,2,\cdots ,N$, respectively, and
experimentally measure $Tr\rho ^{n}$. In this way, we can get $N$ equations
depending on the $N$ eigenvalues. In principle, all the eigenvalues can be
determined by solving these equations. In order to do so, we can define the
generalized swapping operator as $V_{n}\left\vert \psi _{1},\psi _{2},\cdot
\cdot \cdot ,\psi _{n}\right\rangle =\left\vert \psi _{n},\psi _{1},\psi
_{2},\cdot \cdot \cdot ,\psi _{n-1}\right\rangle $. With the swapping
operator one can find $Tr\rho ^{n}=TrV_{n}\rho ^{\otimes n}$. Thus one can
first prepare a probing qubit $\left\vert \varphi \right\rangle _{p}=\frac{1%
}{\sqrt{2}}\left( \left\vert 0\right\rangle +\left\vert 1\right\rangle
\right) $ and $n$ copies of measured state $\rho $. Then let the $n+1$
particles undergo a controlled $V_{n}$ gate, i.e., $\mathbf{1}_{2}\oplus
V_{n}$ with the probing qubit as the control qubit. Finally, the $\sigma _{x}
$ measurement is performed on the probing qubit and the probability of
obtaining $\pm 1$ will be $\frac{1\pm Tr\rho ^{n}}{2}$ which is as expected.
The quantum circuit is shown in Fig. 1. Hence, generally speaking, all the
coherence measures can always be obtained by measuring $N-1$ $V_{n}$ with at
most $N$ copies of the state $\rho $. However, for the coherence measure
based $l_{2}$ norm, one can find that the measurement scheme becomes quite
simple, because it can be directly obtained by only measuring $V_{2}$ with
only 2 copies of $\rho $, which does not depend on the dimension of the
measured density matrix. This is akin to the overlap measurement scheme [16]. 
\begin{figure}[tbp]
\centering
\includegraphics [width=5cm]{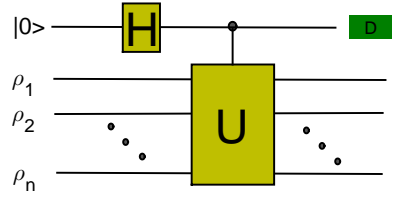}
\caption{ (color online) The circuit for the scheme of measurable total
coherence and the similar probing scheme. This is a generalized QOM with $%
U=V_{n}$. For the quantum probing scheme, the initial state of the probing
qubit is given by $\protect\rho _{p}$ and the initial state $\protect\rho %
_{1}\otimes \protect\rho _{2}\cdots \protect\rho _{n}$ is usually replaced
by some density matrix $\protect\rho _{s}$. In DQC1 scheme, $\protect\rho %
_{s}=\frac{\mathbf{1}_{N}}{N}$ with $N=2^{n}$. }
\end{figure}

\section{The cost of DQC1-like quantum probing schemes}

Here we consider the DQC1-like quantum probing schemes which include the
above mentioned QOM [16] and the remarkable DQC1 schemes [14,15]. The
features of this kind of schemes are (i) a probing qubit is used to extract
the information from the quantum system; (ii) the cost of the probing does
not depend on the probed quantum systems (or the dimension of the input
state space), once the system has been designed. The quantum circuit can be
sketched as Fig. 1, where a probing qubit is sent to the probed quantum
system, and the interaction between the probing qubit and the system is
usually provided by one or several controlled-$U$ operations. It is shown  that there exists quantum speedup in the schemes [14,15], but the essence of this speedup is neither entanglement nor discord between the probing qubit and the probed qubits [14,28,29]. Most people could think that the coherence as the candidate should be an intuitive physics, but no quantitive description has been presented up to now. Here we will do such a job by proving a weak result that nontrivial probing needs the existence of  coherence.

For generality, we set the probing qubit to be given by 
\begin{equation}
\rho _{p}=\frac{1}{2}\left( \mathbf{1}_{2}+\mathbf{P}\cdot \overrightarrow{%
\sigma }\right)
\end{equation}%
where $\mathbf{P}$ is a real 3-dimensional vector with $\left\vert \mathbf{P}%
\right\vert \leq 1$ and $\overrightarrow{\sigma }$ denotes the vector made
up of the 3 Pauli matrices. Suppose that the probed \textit{n}-dimensional
density matrix is denoted by $\rho _{s}$. So the controlled-$U$ operation
will lead to the final state as%
\begin{equation}
\rho _{f}=\left( \mathbf{1}_{n}\oplus U\right) \left( H\rho _{p}H\otimes
\rho _{s}\right) \left( \mathbf{1}_{n}\oplus U^{\dag }\right) ,
\end{equation}%
with $H$ the Hadamard gate. Thus the final density matrix of the probing
qubit becomes%
\begin{equation}
\rho _{pf}=\frac{1}{2}\left( 
\begin{array}{cc}
1+P_{1} & \left( P_{3}+iP_{2}\right) Tr\rho _{s}U^{\dag } \\ 
\left( P_{3}-iP_{2}\right) Tr\rho _{s}U & 1-P_{1}%
\end{array}%
\right) .
\end{equation}%
In order to guarantee that this probing scheme works, it is required that $%
P_{2}$ and $P_{3}$ do not vanish simultaneously. Similarly, if we want to
use $U$ operation to probe information of $\rho _{s}$, $U$ should not be the
identity. Based on Eq. (2) and Eq. (10), one can easily obtain the total
coherence based on l$_{2}$ norm for $\rho _{p}$ and $\rho _{pf}$ as%
\begin{eqnarray}
C\left( \rho _{p}\right) &=&\frac{\left\vert \mathbf{P}\right\vert ^{2}}{2},
\\
C\left( \rho _{pf}\right) &=&\frac{1}{2}\left[ P_{1}^{2}+\left(
P_{2}^{2}+P_{3}^{2}\right) \left\vert Tr\rho _{s}U\right\vert ^{2}\right] .
\end{eqnarray}%
So the change of the total coherence can be given by%
\begin{equation}
\Delta C_{U}=\left\vert C\left( \rho _{pf}\right) -C\left( \rho _{p}\right)
\right\vert =\frac{\left( P_{2}^{2}+P_{3}^{2}\right) }{2}\left( 1-\left\vert
Tr\rho _{s}U\right\vert ^{2}\right) ,
\end{equation}%
with the subscript $U$ denoting the change of the total coherence induced by
the controlled-$U$ operation. Thus $\Delta C_{U}$ is closely related to the
evaluation of this quantum probing scheme. From the point of probing qubit
of view, the cost of the probing qubit is that the total coherence  changes $\Delta C_{U}$  for such a task. Generally, if $U$ is an identity which means we
do nothing in the scheme, one will see that $\Delta C_{U}=0$. For the
general DQC1 where $P_{1}=P_{2}=0$ and $\rho _{s}=\frac{\mathbf{1}_N}{N}$
with $N=2^n$, we have $\Delta C_{U}=\frac{P_{3}^{2}}{2}\left( 1-\left\vert 
\frac{TrU}{2^{n}}\right\vert ^{2}\right) $ which is consistent with Ref. [30].
For the QOM where $P_{1}=P_{2}=0$, $P_{3}=1$ and $\rho_s=\rho_1\otimes\rho_2$%
, one can immediately find that $\Delta C_{V_2}=\frac{1}{2}\left (
1-\left\vert Tr\rho_1\rho_2\right\vert^2\right)$ which is directly given by
the overlap of $\rho_1$ and $\rho_2$.

In fact, the above probing schemes maybe include more unitary operations
denoted by $U_{i}$ (Here we mainly consider the controlled-U operations, and
the unitary operations separately performed on the probing qubit and the
probed quantum state. In particular, it is more reasonable to consider all the operations given by the basic quantum logic gates.). We would like to define the cost of such a scheme $%
\mathcal{C}$ as the sum of the changes of the total coherence of the probing
qubit. That is, 
\begin{equation}
\mathcal{C}=\sum\limits_{k}\Delta C_{k}
\end{equation}%
with $k$ taking all the unitary operations. In particular, one should note
that $\Delta C_{k}\geq 0$ for any $k$ based on Eq. (15). Therefore,
generally speaking, the more operations are used, the more cost is paid. In
particular, $\mathcal{C}$ will not vanish if the probing scheme only
includes two unitary operations such as $U$ and $U^{\dagger }$. This should
be distinguished from the scheme which only includes a single identity
operation. This difference can be understood in the frame of basic logic gates.
That is, suppose $U$ and $U^\dagger$ are given by a series of logic gates, the qubits through them
will have to undergo the corresponding `dynamical evolution', even though at the final moment the original state is recovered. On the contrary, a direct identity operation (we mean no logic gates), no such an `evolution' is needed. In this sense, the subscript $k$ in Eq. (15) taking all the covered logic gates could
be more reasonable, but it will lead to more complicated calculations because how to construct a given
operation by logic gates has to be considered.

Finally, one can find that the coherence measures given by Eqs. (3,4) are described by the
eigenvalues of the density matrix. In the above probing scheme, one can
easily calculate that the eigenvalues of the density matrix $\rho _{pf}$ are
the functions of $\left( P_{2}^{2}+P_{3}^{2}\right) \left\vert Tr\rho
_{s}U\right\vert ^{2}$. Hence we can always find what the cost $\mathcal{C}$
is for the different measures which we can choose. In particular, it can be
shown that $\mathcal{C}$ is directly related to $\left(
P_{2}^{2}+P_{3}^{2}\right) \left\vert Tr\rho_sU\right\vert ^{2}$. To sum up, 
$\mathcal{C}$ vanishes if and only if the probing scheme is trivial. In this
sense, we think that $\mathcal{C}$ can be understood as the cost of such a
probing scheme.

\section{Discussion and conclusion}

We have studied the total coherence of a quantum state and presented several
coherence measures which are independent of the basis. It is shown that all
the presented measures especially including the measure based on $l_2$ norm
satisfy all properties such as the monotonicity. This actually provides a
very convenient tool for the relevant researches due to the simple form of $%
l_2$ norm. In particular, we have shown that the total coherence measures based on the relative entropy and the $l_2$ norm have the same expression by optimizing the basis or by quantifying the distance. 
 In addition, for integrity, the experimental schemes for the
detection of coherence are also briefly introduced. 
Finally, we study the total coherence in the DQC1-like quantum probing
schemes. It is shown that both the normalized trace in DQC1 and the overlap
of the two states in QOM can be well described by the change of the total
coherence of the probing qubit. In other words, all the nontrivial probing
schemes have to lead to the change of the total coherence. Therefore, this
change can be understood as the cost of implementing such a probing scheme.
This could motivate a new platform to study the essence of the speedup of
mixed-state quantum computing.

\end{document}